\documentclass{article}

\usepackage[utf8]{inputenc}
\usepackage{subcaption}
\usepackage{geometry}
\usepackage{authblk}
\usepackage{amsmath}
\usepackage{amssymb}
\usepackage{upgreek} % for upright greek letters
\usepackage{graphicx}
\usepackage[numbers,sort&compress]{natbib}
\usepackage{hyperref}
\usepackage{siunitx}
\DeclareSIUnit\angstrom{\text{\AA}}
\usepackage{caption}
\usepackage{xcolor}
\usepackage{setspace}
\usepackage{titlesec}
\usepackage{enumitem}
\usepackage{quoting}
\usepackage{etoolbox}
\usepackage{epigraph}
\usepackage{multicol}
\usepackage{float}
\usepackage{miller}
\usepackage[percent]{overpic}
\usepackage{gensymb}
\titleformat{\section}
  {\color{black}\normalfont\Large\bfseries}
  {}{0pt}{}
\titlespacing*{\section}{0pt}{1.5em}{0.5em}
\titleformat{\subsection}
  {\color{black}\normalfont\large\bfseries}
  {}{0pt}{}
\titlespacing*{\subsection}{0pt}{1.2em}{0.3em}

\hypersetup{
  colorlinks=true,
  linkcolor=black,
  urlcolor=black,
  citecolor=black
}

\makeatletter
\renewcommand{\maketitle}{%
  \begin{center}
{\Large\bfseries\@title\par}%
\end{center}
  \vskip 1em%
  \begin{center}%
    {\normalsize\@author}%
    \vskip 1em%
    {\normalsize\@date}%
  \end{center}%
}
\makeatother

\title{Microscopic and macroscopic characterization:\\ MBE-grown versus sputter-deposited Au/Co/Au thin films\\ 
for CISS and MIPAC effect studies}

\author[1,3]{Lokesh Rasabathina}
\author[1]{Thi Ngoc Ha Nguyen}
\author[5]{Aleksandr Kazimir}
\author[1,3]{Rico Ehrler}
\author[1,3]{Julia Krone}
\author[1,3]{Franziska Sch\"olzel}
\author[1,3]{Zihao Liu}
\author[1,3,4]{Peter Heinig}
\author[2]{Markus G\"o\ss ler}
\author[6]{Irene Coin}
\author[5]{Christina Lamers}
\author[1,3]{Georgeta Salvan}
\author[7]{Lech Tomasz Baczewski}
\author[1]{Christoph Tegenkamp}
\author[1,3,4]{Olav Hellwig}

\affil[1]{Institute of Physics, Chemnitz University of Technology, Chemnitz, Germany}
\affil[2]{Institute of Chemistry, Chemnitz University of Technology, Chemnitz, Germany}
\affil[3]{Center for Materials, Architectures and Integration of Nanomembranes (MAIN), Chemnitz University of Technology, Chemnitz, Germany}
\affil[4]{Helmholtz-Zentrum Dresden-Rossendorf, Dresden, Germany}
\affil[5]{Institute for Drug Discovery, Leipzig University, Leipzig, Germany}
\affil[6]{Institute of Biochemistry, Leipzig University, Leipzig, Germany}
\affil[7]{Institute of Physics, Polish Academy of Sciences, Warszawa, Poland}

\date{}

\begin{document}

\maketitle

\vspace{1em}
\noindent\rule{\textwidth}{0.4pt}
\vspace{0.5em}
\noindent
{\bfseries Abstract}
\vspace{0.5em}

Chirality-induced spin selectivity (CISS) enables spin-dependent transport at chiral molecule/Au\hkl(111) interfaces and is used in spintronics when combined with ferromagnetic thin films in spin-valve-type hybrids. However, the influence of substrate microstructure on CISS and the related magnetization induced by the proximity of adsorbed chiral molecules (MIPAC) effect is still not well understood. In this study, we compare the effects of the adsorption of L-chiral $\alpha$-helical alanine-rich peptides on Au/Co/Au ferromagnetic thin films fabricated by molecular beam epitaxy (MBE) and magnetron sputtering. X-ray reflectivity and X-ray diffraction show sharper interfaces and a narrower Au\hkl(111) rocking-curve width for the MBE-grown sample. However, atomic force microscopy and scanning tunneling microscopy images reveal that both sample types have locally smooth Au\hkl(111) surface regions suitable for peptide adsorption, despite clear differences in larger-scale morphology. Microscopic scanning tunneling spectroscopy after peptide exposure yields similar magnetization-direction-dependent tunneling currents in both sample types, confirming a similar magnitude CISS effect on the molecular scale. In contrast, macroscopic magneto-optical Kerr effect hysteresis loops and effect microscopy reveals that only sputter-deposited samples show slight coercivity enhancements and a consistent reduction in domain wall velocity after peptide exposure. These results suggest that microscopic CISS signatures are robust for both sample types, whereas macroscopic MIPAC-type magnetic responses are more sensitive to the substrate microstructure.

\vspace{1em}
\noindent
\textbf{Keywords:} CISS, MIPAC, sputter deposition, MBE, chiral molecules, peptides, magnetic thin films, STM, STS, AFM, XRD, XRR, MOKE

\vspace{0.5em}
\noindent\rule{\textwidth}{0.4pt}
\vspace{1.5em}

\section{1 Introduction}

The chirality-induced spin selectivity (CISS) effect is a quantum phenomenon, where chiral molecules, depending on their left- or right-handed chirality, let electrons of a certain spin direction preferentially pass through, essentially acting as an interfacial spin filter \cite{rayAsymmetricScatteringPolarized1999,naamanChiralInducedSpinSelectivity2012,bloomChiralInducedSpin2024, xieSpinSpecificElectron2011,weiMolecularChiralityCharge2006,rayChiralityInducedSpinSelectiveProperties2006,aragonesMeasuringSpinPolarizationPower2017}. Over the last years, CISS has been observed in several molecular and hybrid systems, including helical peptides, DNA, helicenes \cite{safariDepositionChiralHeptahelicene2022,kettnerChiralityDependentElectronSpin2018,safariSpinSelectiveElectronTransport2024} and other chiral organic structures, showing that molecular chirality can be directly connected to spin-dependent electron transport. However, the microscopic mechanism behind this effect is still under discussion, especially with respect to whether the spin selectivity mainly originates from transport through the helical molecule, from the molecule/substrate interface, or from the combined hybrid system consisting of the molecule/substrate interface and an additional ferromagnetic layer below. In molecule-ferromagnetic thin film hybrid systems, the CISS effect can be observed by measuring the changes in the tunneling current upon reversing the magnetization direction of the ferromagnetic layer using scanning tunneling microscopy/spectroscopy (STM/STS) \cite{nguyenCooperativeEffectElectron2022,nguyenMechanismMolecularCISS2024}. In this type of experiment, the magnetic thin film usually has out-of-plane uniaxial anisotropy and thus provides stable switchable states with the magnetization up or down with respect to the surface, while the adsorbed chiral molecules form the spin-selective interface via adsorption onto the Au\hkl(111) surface. Therefore, a magnetization-direction-dependent tunneling current can be used as a microscopic signature of a spin-valve-type response in the molecule/ferromagnetic thin film hybrid structure.
\\

On the other hand, it has also been reported that the chiral molecules adsorbed on ferromagnetic thin films can alter the magnetic properties of the thin film, an effect which is referred to as magnetization induced by proximity of adsorbed chiral molecules (MIPAC) \cite{sharmaControlMagnetoopticalProperties2020,bendorMagnetizationSwitchingFerromagnets2017}. Such MIPAC-type effects are usually detected through macroscopic magnetometry measurements, for example, by changes in coercivity or a hysteresis loop shift similar as observed for exchange-bias systems \cite{raduExchangeBiasEffect2008,noguesExchangeBias1999,maatPerpendicularExchangeBias2001}. Thus, while STS mainly probes the local microscopic electronic response of the molecule/substrate interface, MIPAC-type measurements probe the magnetic response of the ferromagnetic thin film at a larger scale. This interplay between the chiral molecules and magnetic thin films has potential applications for spintronic sensors and devices \cite{matsuzakaMagnetoresistanceEffectBased2025,mathewNonmagneticOrganicInorganic2014,firouzehChiralityInducedSpinSelectivity2024}. For this reason, it is important to understand not only whether a CISS-related signal can be detected, but also how the magnetic substrate itself influences the measured microscopic and macroscopic responses. 
\\

While the CISS effect has been widely observed in several helical or chiral molecule-thin film systems, the majority of those thin films are fabricated using molecular beam epitaxy (MBE) \cite{nguyenMechanismMolecularCISS2024,banerjee-ghoshSeparationEnantiomersTheir2018,zivAFMBasedSpinExchangeMicroscopy2019,ranaChiralityInducedSpinSelectivity2025}, which leads to the films having atomically smooth surfaces and minimal interlayer roughness and intermixing \cite{linNanometerthickMolecularBeam2024,ranaEnantiospecificMagnetoconductanceAsymmetry2026}. MBE-grown Au/Co/Au substrates are therefore very useful for fundamental CISS studies, because the Au capping layer protects the ultrathin Co layer from oxidation and provides a suitable, well-studied Au\hkl(111) surface for molecular adsorption. The smooth surface and sharp interfaces also help to reduce structural uncertainty when correlating molecular adsorption with local tunneling spectroscopy. At the same time, such well-defined epitaxial substrates do not represent all magnetic thin-film microstructures that may be relevant for future device-oriented studies. The goal of this work is to extend CISS and MIPAC effect experiments also to non-epitaxial sputter-deposited ferromagnetic thin film systems, as those allow more versatile layer structures, a more application-friendly fabrication \cite{alfonsoThinFilmGrowth2012} and avoid the use of expensive single crystal substrates. Magnetron sputtering is widely used for metallic multilayers and allows flexible control of layer thicknesses, seed layers, cap layers, and substrate choice. However, non-epitaxial sputter-deposited films can differ strongly from epitaxial MBE-grown films in their interface roughness, crystallite orientation distribution, grain structure, and larger-scale surface morphology. These structural differences may influence peptide adsorption, local tunneling conditions, magnetic anisotropy, and magnetic domain-wall motion. Therefore, a direct comparison between MBE-grown and sputter-deposited Au/Co/Au substrates is essential to understand whether microscopic CISS signatures and macroscopic MIPAC-type magnetic responses are affected in the same way by substrate microstructure. 
In the current study, we compare the microscopic and macroscopic signatures of CISS and MIPAC on Au/Co/Au samples fabricated by both MBE and magnetron sputtering, before and after exposure to $\alpha$-helical alanine-rich peptides (L-polyalanine, LPA). The two substrate types are first carefully characterized by X-ray reflectivity, X-ray diffraction, atomic force microscopy, and scanning tunneling microscopy, in order to identify differences in the buried Co-layer structure as well as at the local Au\hkl(111) surface. This is important because XRR and XRD probe interface quality, intermixing, and Au\hkl(111) crystallite alignment, whereas AFM and STM provide real-space information about the surface morphology that is directly relevant for peptide adsorption. By correlating structural properties with STS and magnetic field reversal measurements, we aim to better understand the role of the magnetic thin film's microstructure in the observation of features related to the CISS and MIPAC effects, such as changes in coercivity or exchange-biased field reversal curves \cite{raduExchangeBiasEffect2008,noguesExchangeBias1999,maatPerpendicularExchangeBias2001}. The microscopic CISS-related response is studied by magnetisation-dependent STS measurements after peptide adsorption, while the macroscopic magnetic response is investigated by MOKE hysteresis loops and domain-wall motion measurements before and after peptide exposure. A careful and comprehensive comparison of MBE-grown versus sputter-deposited samples allows us to distinguish between local CISS signatures measured by STS and macroscopic magnetic changes measured by integrated MOKE hysteresis loops and microscopy.

\section{2 Experimental methods}

\subsection{2.1 Sample fabrication and layer-stack design}
Both MBE-grown and sputter-deposited samples consist of a Pt/Au/Co/Au layer stack. The MBE samples were grown on sapphire c-plane (Al$_2$O$_3$(0001)) epi-ready substrates, while the sputter-deposited samples were fabricated on thermally oxidized Si substrates with 100\,nm SiO$_2$.  Since Au\hkl(111) is a highly studied surface for molecular adsorption \cite{santiago-rodriguezAtomicMolecularAdsorption2014}, it is commonly used as a cap layer on top of the FM layer \cite{nguyenCooperativeEffectElectron2022,nguyenMechanismMolecularCISS2024,giaconiEfficientSpinSelectiveElectron2023} and also serves as the seed layer. A thin Co layer with out-of-plane magnetic anisotropy is often used as the ferromagnetic layer \cite{jomniFaceCentredCubic2000}. The Pt layer below the bottom Au seed layer serves as an epitaxial bridge layer between the single crystal Al$_2$O$_3$(0001) substrate and the Au \cite{farrowEpitaxialGrowthPt1993} in order to finally obtain a Au\hkl(111) surface in the MBE-grown samples. For the sputter-deposited samples, the Pt layer is used to induce an fcc\hkl(111) out-of-plane texture on the amorphous Ta adhesion layer deposited on the Si/SiO$_2$ wafer substrate \cite{ehrlerManifoldRoleTa2025}, thereby also promoting a Au\hkl(111) out-of-plane orientation at the surface. For the epitaxial MBE-grown samples the Co thickness window for achieving perpendicular magnetic anisotropy is significantly larger (\SIrange{0.5}{2}{\nano\meter}) (refer to Fig. SM1 in supplementary material of \cite{bendorMagnetizationSwitchingFerromagnets2017}) due to less interfacial roughness and reduced interdiffusion as compared to the sputter-deposited samples (0.5 - 1.1\,nm) (see Fig. S1 in supplementary material), so a larger Co layer thickness of 1.2 nm is chosen for the MBE-grown sample, while for the sputter-deposited sample a Co layer thickness of only 0.9 nm is chosen.

\vspace{0.5em}

\subsubsection{MBE-grown (epitaxial) Au/Co/Au substrates}
This sample was fabricated using MBE. A \SI{8}{\milli\meter} $\times$ \SI{5}{\milli\meter} Al$_2$O$_3$(0001) substrate was used and the layer stack consists of Pt (5\,nm) / Au (20\,nm) / Co (1.2\,nm) / Au (5\,nm) as shown in Fig. 1(a). The Pt layer was deposited at \SI{700}{\celsius} and the substrate was cooled down to RT. Next, a 20\,nm Au seed layer was deposited at RT and subsequently annealed at \SI{250}{\celsius} for 2 hours. Knudsen effusion cells were used for Au and Co deposition, while Pt was deposited using an electron gun. Similar samples were used previously to study the CISS effect in hybrid nanostructures composed of chiral molecules adsorbed on the Au top layer surface \cite{nguyenCooperativeEffectElectron2022,zivAFMBasedSpinExchangeMicroscopy2019,kaponControllingAmyloidAssembly2025,rohmerChiralInducedSpin2025}. 

\subsubsection{Sputter-deposited Au/Co/Au substrates}
This sample was fabricated using a confocal sputter-up geometry of an AJA International ATC magnetron sputtering system. A \SI{14}{\milli\meter} $\times$ \SI{14}{\milli\meter} thermally oxidized Si/SiO$_2$ (100\,nm) substrate was used and the layer stack consists of Ta (2\,nm) / Pt (5\,nm) / Au (20\,nm) / Co (0.9\,nm) / Au (5\,nm) as shown in Fig. 1(b). The deposition parameters for each layer are as follows: Ta (2\,nm) was deposited at 3\,mTorr Ar pressure, 50\,W, \SI{0.431}{\angstrom/\second} (DC mode); Pt (5\,nm) at 1.5\,mTorr Ar pressure, 60\,W, \SI{0.911}{\angstrom/\second} (DC mode); Au seed (20\,nm) and Au cap (5\,nm) at 1\,mTorr Ar pressure, 60\,W, \SI{0.959}{\angstrom/\second} (RF mode); Co (0.9\,nm) at 1\,mTorr Ar pressure, 120\,W, \SI{0.486}{\angstrom/\second} (RF mode). All depositions were performed at room temperature. The AJA tool provides a base pressure below $5 \times 10^{-8}$\,Torr, with substrate rotation at 75 rpm to ensure uniform film thickness. After deposition, samples were stored in a nitrogen environment to minimize contamination. The samples were cut into \SI{8}{\milli\meter} $\times$ \SI{5}{\milli\meter} sizes for the measurements. For X-ray photoelectron spectroscopy (XPS) measurements, thermally oxidized Si/SiO$_2$ (100\,nm) substrate was used and the layer stack consists of Ta (2\,nm) / Au (25\,nm), where Ta (2\,nm) was deposited at 3\,mTorr Ar pressure, 50\,W, \SI{0.431}{\angstrom/\second} (DC mode) and Au (25\,nm) at 1\,mTorr Ar pressure, 60\,W, \SI{0.959}{\angstrom/\second} (RF mode).

\subsection{2.2 Peptide preparation and deposition}
Alanine-rich peptides consisting of alanine [A], lysine [K] and cysteine [C], having the general formula $\mathrm{C[AAAAK]_n}$, were used in this study. Cysteine, containing a thiol group, is used to form the Au-S bond at the Au surface \cite{loveSelfAssembledMonolayersThiolates2005}, while alanine induces a helical structure. Lysine is added to the sequence to improve solubility \cite{marquseeHelixStabilizationGluLys1987} in polar solvents. For the STM/STS and MOKE measurements, a commercially purchased 36-mer (n=7) was used. The peptides were dissolved in absolute ethanol to obtain a concentration of 1\,mM. The substrates were cleaned with absolute ethanol and dried thoroughly with nitrogen and immediately dipped into the peptide solution of 1\,mM concentration for 10 hours before doing the MOKE experiments. After 10 hours, the substrates were taken out of the peptide solution, washed with absolute ethanol to remove the non-adsorbed peptide layers and dried with nitrogen gas. For the STM/STS measurements, the peptides were diluted to 0.1\,mM concentration and drop-cast on the samples. It must be noted that the commercially purchased peptides were used for STM/STS and MOKE measurements. For the X-ray photoelectron spectroscopy (XPS), in-house synthesized 11-mer (n=2), 16-mer (n=3) were used along with the commercially purchased 36-mer, at a concentration of 1\,mM. A more detailed synthesis procedure employed for in-house synthesis can be found in \cite{kazimirAlanineRichPeptideSpinDependent2026}.

\subsection{2.3  Characterization methods}
The characterization methods used in this study were applied to the samples depending on the aim of each measurement. When a technique was used before and after peptide exposure, the same measurement parameters were maintained for both the MBE-grown and sputter-deposited samples to allow a direct comparison.

\subsubsection{X-ray reflectivity/diffraction (XRR/XRD) }
X-ray reflectivity (XRR) and X-ray diffraction (XRD) were performed before peptide exposure, utilizing a rotating-anode Rigaku SmartLab diffractometer in parallel beam configuration with a 5-axis goniometer with an OOP movable source arm, an OOP ($\theta$) and IP ($\theta_\chi$) movable detector arm, sample IP rotation ($\phi$), and sample tilt ($\chi$) perpendicular to the plane of beam direction and film normal.

XRR was aligned to the physical surface of the samples, ignoring any crystalline alignment effects (such as substrate miscuts). A K$_\upbeta$ filter was used to balance beam intensity and spectral resolution. The resulting XRR curves were fitted using GenX3, version 3.6.27 [28] to extract layer thicknesses, interface roughnesses, and the scattering length density (SLD) depth profile. The density profile was modeled using the following stacks:
\begin{itemize}
    \item MBE: Al$_2$O$_3$(substrate) / Pt / Au$ _{\text{seed}}$ / Co / Au $_{\text{cap}}$
    \item Sputter: SiO$ _{\text{2}}$(substrate) / Ta / Pt / Au$ _{\text{seed}}$ / Co / Au $_{\text{cap}}$
\end{itemize}
Thickness and roughness were fitted for all layers. Density was refined only for the Co layers, with the other layers held at their bulk values. A Gaussian beam footprint was used to account for behavior in the total-reflection regime. Resolution effects were fitted using the "fast conv" resolution model. 

Symmetric out-of-plane XRD (OOP-XRD) was measured using a two-bounce Ge(220) monochromator, selecting the Cu K$_{\upalpha,1}$ line. The MBE sample was aligned to the substrate normal to account for miscut. This was not necessary for the sputter-deposited sample, as amorphous SiO$_2$ and Ta layers are present between the crystalline Si and the film, preventing epitaxy. The sputter-deposited sample was instead aligned to the physical sample surface, analogous to XRR. 
Rocking scans were performed using the same optics as the OOP-XRD scans, maintaining a constant $2\theta$ angle and varying the incidence angle $\omega$. The Rocking angle $\omega$ denotes, in our case, the angle of incidence of the X-ray beam relative to the sample surface.

In-plane XRD (IP-XRD) was measured utilizing a goniometer axis that moves the detector in the film plane ($\theta_\chi$). To achieve a symmetric scan, the sample and detector were rotated in a 1:2 ratio, yielding a $\phi-2\theta _{\chi} $ scan.
The samples were aligned to the IP Au\hkl(-220) peak. The OOP detector angle is fixed at 0\degree, and the incidence source angle was optimized for maximum intensity: $\omega = \SI{0.58}{\degree}$ for the MBE and $\omega = \SI{0.55}{\degree}$ for the sputter-deposited sample. No filter or monochromator was used to maximize intensity.
Additionally, $\phi$ scans were performed for the IP Au\hkl(-220) peak with the same alignment and optics as the IP-XRD scan, holding $2\theta _{\chi}$ constant and varying the sample angle $\phi$.

Pole figures were performed using the same geometry as the IP-XRD measurements. The angle of incidence $\omega=\theta$ was held constant, and the detector moved from the IP to the OOP direction using the 2-axis movable IP detector arm, maintaining an angle of $2\theta$ relative to the incidence beam. In this case, the angle between the detector and the film normal is the polar angle $\alpha$. At each $\alpha$, the sample was rotated one full $\phi$ revolution, measuring the azimuthal angle $\beta$. The data is plotted in spherical coordinates, with data interpolation between the measured points.

\subsubsection[STM/STS and Delta z measurements]{Atomic force microscopy (AFM), scanning tunneling microscopy (STM), scanning tunneling spectroscopy (STS) and magnetization-dependent $\Delta z$ measurements}
AFM and STM measurements were performed on both sample types to characterize the surface topography (before peptide exposure), while the STS measurements were performed to measure current-voltage (I-V) characteristics (after peptide exposure). AFM was measured using an AIST-NT SmartSPMTM 1000 setup. The AFM measurements were carried out under ambient conditions using a NSG10 tip, which is commercially purchased with a typical tip radius of approximately 6\,nm. STM/STS measurements were performed under ambient conditions at room temperature. A Pt tip was used and the measurements were done in constant current mode, i.e., the tip-sample distance ($\Delta$z) was auto-adjusted to keep the tunnel current constant. First, STM imaging was performed to record the surface topography of the thin films. Then STS was performed to measure I-V curves at various spots on both MBE-grown and sputter-deposited samples after exposure to the peptides. For each sample, many I–V curves were recorded with the Co layer magnetized in the up direction and again after switching the magnetization to the down direction using an external magnet. All the curves measured over various spots on the sample surface were then averaged. Using the same setup, magnetization-dependent tip-sample distance measurements were performed. In this method, the feedback loop continuously adjusts the tip's vertical position to maintain a constant tunneling current. An external magnet is brought near the sample to remagnetize the Co layer. The magnet is flipped repeatedly to switch the magnetization direction of the Co layer between up and down magnetization. The tip position is recorded relative to the magnetization direction of the Co layer to monitor any changes in magnetoresistance.  

\subsubsection{X-ray photoelectron spectroscopy (XPS)}
The X-ray photoelectron spectroscopy (XPS) analysis was performed with an ESCALAB 250Xi photoelectron spectrometer (Thermo Fisher Scientific) in an ultra-high vacuum (UHV) chamber using a monochromatic Al K$_\alpha$ (\SI{1486.68}{\electronvolt}) X-ray source and a beam diameter of \SI{650}{\micro\meter}. The base pressure was approximately $2 \times 10^{-9}$\,mbar. The binding energies of all spectra are referenced to the binding energy of Au 4f$_{7/2}$ (84 $\pm$ 0.1~eV). Survey spectra were collected with a pass energy of \SI{200}{\electronvolt}, while core-level spectra used \SI{20}{\electronvolt}. For the measurements, four Au substrates were prepared: one absolute ethanol-treated reference substrate and three peptide-exposed substrates with 11-mer, 16-mer, and 36-mer peptides, respectively. 

\subsubsection{Magnetometry measurements: Magneto-Optical Kerr Effect (MOKE) hysteresis loops/ microscopy}
MOKE microscopy was performed using an Evico Magnetics system equipped with a 20$\times$ objective. Out-of-plane (OOP) magnetic fields were applied using an electromagnet. Hysteresis loops were measured with polar sensitivity between -300 and +300\,Oe with a step size of 5\,Oe and a step delay of 0.5\,s. To investigate domain dynamics, a ``pulsing domains'' protocol was employed: first, the switching field was identified by hysteresis measurement (e.g., +130\,Oe for sputter-deposited sample). The sample was saturated at -300\,Oe, then the field was increased to just below the switching field (e.g., +120\,Oe for the sputter-deposited sample) until domain nucleation was observed due to the thermal excitations at room temperature. At this point, the field was turned off, and short magnetic field pulses (+120\,Oe, 100\,ms each, total 500\,ms) were applied. Images were captured before and after 500\,ms pulsing, and the resulting movement of the magnetic domain wall was quantified by overlaying the images and extracting the distance moved by the domain wall at 10 random points, perpendicular to the domain wall motion direction. This procedure was performed both before and after peptide exposure. The nucleation field and the pulsing field values taken before peptide exposure were also used for nucleation and pulsing after peptide exposure. It must be noted that after the samples were taken out of the peptide solution, they were washed with absolute ethanol to remove the non-adsorbed peptides and dried with nitrogen gas and the MOKE measurements were performed immediately afterward. \\

\section{3 Results and discussion}

\subsection{3.1 Structural comparison (XRR/SLD/out-of-plane and in-plane XRD, AFM and STM)}

\begin{figure}[H]
    
    \includegraphics[width=\textwidth]{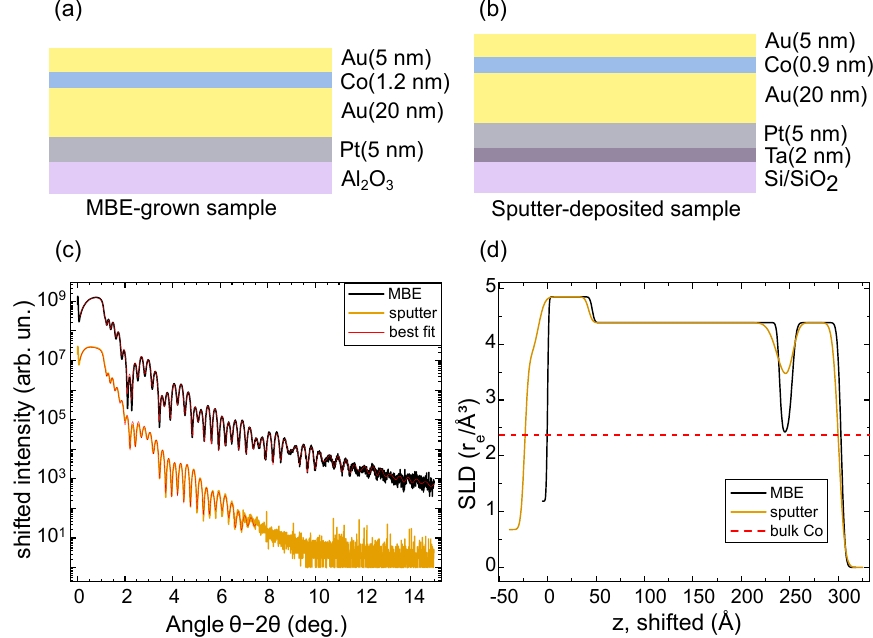}
    \caption{ (a,b) layer stack of MBE-grown sample and sputter-deposited sample; (c) XRR curves and corresponding best fits (red lines);
(d) Scattering length density (SLD) profiles derived from XRR data, the depth $z$ is aligned to the center of the Co film, with the literature value of bulk Co in red.}
    \label{fig:xray_combined_2}
\end{figure}

The XRR curves for both samples show characteristic thickness oscillations arising from the different electron density profiles of the layers. The MBE-grown sample maintains pronounced oscillations at higher angles (see Fig. 1(c)), indicating smoother interfaces than the sputter-deposited sample. The scattering length density (SLD) profiles (see Fig. 1(d)), extracted from the best fits, confirm the interface quality of both samples. The MBE-grown sample displays sharp, step-like changes at each interface, indicating minimal intermixing or roughness. On the other hand, the sputter-deposited sample shows broader, more gradual transitions, suggesting a much higher interface roughness and partial intermixing between the layers. Furthermore, the MBE-grown 1.2 nm Co layer reaches almost the Co bulk SLD value ($\rho_{\mathrm{Co}} \approx 2.46~r_e/\text{\AA}^3$), while the sputter-deposited 0.9\,nm Co layer has its minimum value far from that, which is again a sign of interfacial intermixing between Co and Au or increased buckling of the Co layers sandwiched between the Au seed and cap layer.
\\

\begin{figure}[H]\centering
    
    \includegraphics[width=\textwidth]{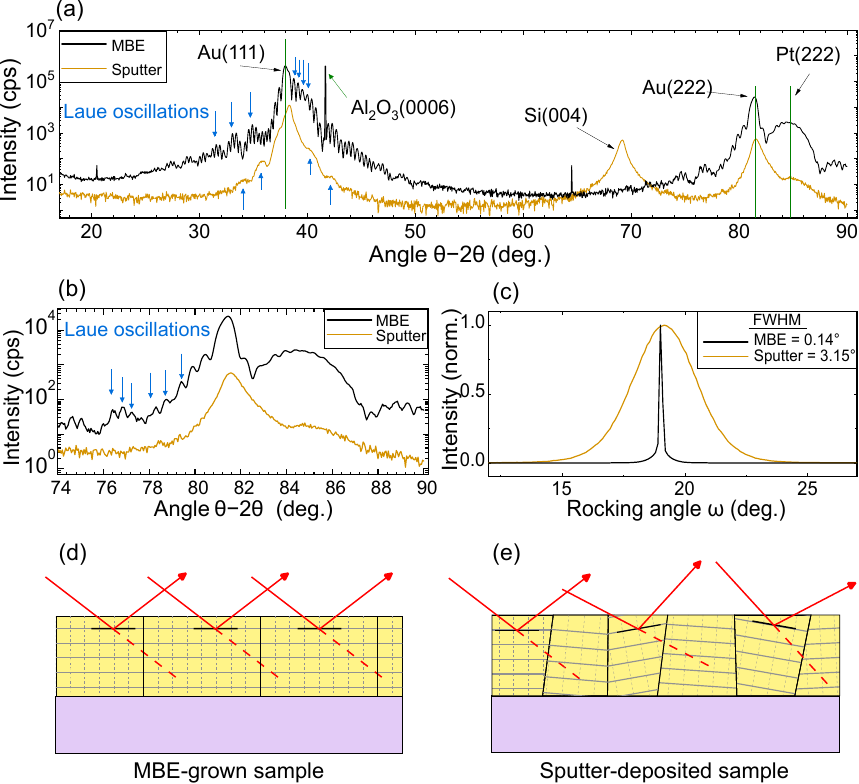}
    \caption{(a) XRD patterns of MBE-grown and sputter-deposited Au/Co/Au samples highlighting sharper peaks and intense Laue oscillations for the MBE-grown sample, which are way less pronounced for the sputter-deposited sample; (b) Enlarged view of XRD patterns in the 2nd order to better separate the Au\hkl(222) and Pt\hkl(222) peak;  (c) Rocking curves measured at the Au\hkl(111) reflection, with FWHM of \SI{0.14}{\degree} and \SI{3.15}{\degree} for MBE-grown and sputter-deposited samples, respectively; (d,e) Schematic representation (not to scale) comparing the more uniform out-of-plane crystallite alignment in the MBE-grown sample with the broader crystallite-orientation distribution in the sputter-deposited sample, consistent with the pronounced Laue oscillations and narrow rocking curve of the MBE-grown sample and the weaker Laue oscillations and broader rocking curve of the sputter-deposited sample.}
    \label{fig:xray_combined_1}
\end{figure}

XRD measurements reveal that both samples display a pure OOP fcc\hkl(111) texture, with characteristic Au\hkl(111) and Au\hkl(222) peaks (see Fig. 2(a)). The MBE-grown sample displays a massively higher peak intensity (note the log scale) compared to the sputter-deposited sample, combined with much more pronounced long-range Laue oscillations (see Fig. 2(a,b)), indicating significantly higher crystalline quality and smoother interfaces as expected for an epitaxial MBE-grown sample \cite{millerExtractingInformationXray2022}. In contrast, the sputter-deposited sample shows less-intense Bragg peaks and less-pronounced Laue oscillations, indicating lower crystallinity and increased interface roughness. Rocking scans were performed to further analyze the out-of-plane crystallite alignment. The full width at half maximum (FWHM) for the MBE-grown sample is \SI{0.14}{\degree}, indicating much more uniform out-of-plane alignment of the crystallographic Au\hkl(111) planes compared to the sputter-deposited sample with a FWHM of \SI{3.15}{\degree}. These results can now be explained using the illustrations in Fig. 2(d,e). The good out-of-plane crystallite alignment for the MBE-grown sample, combined with the low roughness at the layer boundaries, results in strong interference, yielding long-range Laue oscillations. On the other hand, the crystallites for the sputter-deposited sample barely interfere due to the high Rocking width. Combined with the larger layer roughness determined from XRR, the Laue oscillations are short-ranged and quickly damped out. Note that the Laue oscillations of both samples show very different crystal thicknesses due to the difference in coherence.

\begin{figure}[H]\centering
    \includegraphics[width=0.9\textwidth]{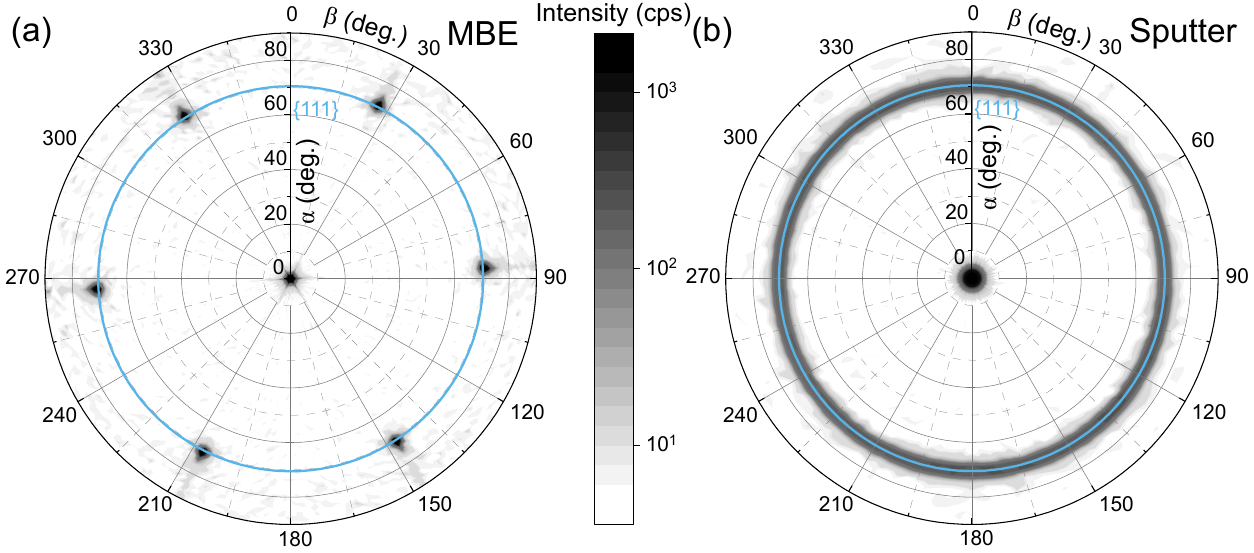}
        \caption{Pole figure measurements for the Au\hkl{111} lattice plane family for (a) the MBE and (b) the sputter-deposited sample. $\alpha$ denotes the polar angle relative to the film normal, and $\beta$ is the azimuthal angle of the IP rotation. The pole figures were measured at the $2 \theta$ angles extracted from the OOP-XRD in Fig. 2(a) of $37.92\degree$ and $ 38.33\degree$ for the MBE and sputter-deposited samples, respectively. The blue line indicates a value of $\alpha=70.53\degree$, at which the \hkl{-111} reflexes should appear. }
    \label{pole_figure}
\end{figure}

To further analyze the texture of the films, we measured pole figures at the angles corresponding to the Au\hkl(111) reflections. In the pole figure, $\alpha$ denotes polar angle relative to the film normal, and $\beta$ is the azimuthal angle of the IP rotation, measured at a constant $2 \theta$ angle, probing the angular distribution of a certain lattice spacing. We selected \SI{37.92}{\degree} for the MBE sample in Fig. 3(a) and \SI{38.33}{\degree} for the sputter-deposited sample in Fig. 3(b), as extracted from Fig. 2(a).

For a purely fcc\hkl(111) textured film, four distinct spots should be present, one at the center ($\alpha=0\degree$) and three at $\alpha=70.53 \degree$, rotated each by $\Delta \beta = 120 \degree$, giving a 3-fold symmetry.
The MBE sample in Fig. 3(a) displays a center spot with six additional peaks at $\alpha \approx 70 \degree$, which are $\Delta \beta =60\degree$ apart, resulting in a six-fold symmetry. This indicates that two competing fcc stacking directions, namely ABC and CBA, are present in the film. Grains with these stacking nucleate on the substrate, forming distinct grain boundaries since they cannot merge. This results in doubling of the symmetry, as the ABC and CBA directions are rotated by $30\degree$.
The sputter-deposited sample, on the other hand, shows a broader center peak with a continuous ring at $\alpha=70.53\degree$. This is a clear sign of a fiber texture, with a well-defined order in the OOP direction but a polycrystalline orientation in the plane, resulting in all directions of the \hkl{-111} equivalent planes being present in the sample, superimposing the three spots per grain into a ring for the full sample.

\begin{figure}[H]\centering
    \includegraphics[width=\textwidth]{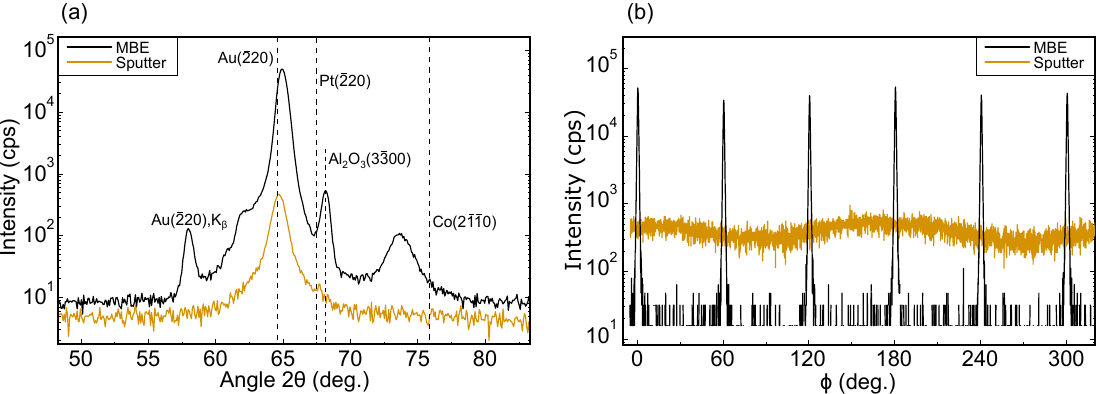}
        \caption{(a) XRD in-plane Bragg scans indicating the Au\hkl(-220) peak from the MBE-grown and sputter-deposited sample. We can also see Pt\hkl(-220), Co\hkl(2-1-10) and K$_\beta$ Au\hkl(-220)(arising from K$_\beta$ radiation) peaks only from the MBE-grown sample; (b) in-plane $\phi$-scans of the Au\hkl(-220) Bragg peak. The MBE sample shows the expected 6-fold symmetry of the fcc\hkl(-220) lattice family, whereas the sputter-deposited sample has an almost constant intensity independent of $\phi$, indicative of a polycrystalline distribution. The intensity variations are caused by the beam footprint on the samples.}
    \label{xrd_inplane}
\end{figure}

To verify the epitaxial quality of the MBE sample, we performed IP-XRD scans, which are displayed in Fig. 4(a). The MBE sample shows 3 pronounced crystal phases, corresponding to the Au\hkl(-220), Pt\hkl(-220) or Al$_2$O$_3$\hkl(3-300), and Co\hkl(2-1-10), which are the planes perpendicular to the fcc Au\hkl(111) and Pt\hkl(111), as well as the hcp Co\hkl(0001) in the OOP direction. Since we did not use a filter, the X-ray spectrum of the tube includes a parasitic K$_\beta$ line from Au\hkl(-220). The Au\hkl(-220) peak shape is also due to the tube spectrum and is not related to the crystal structure. The other peaks are not intense enough to generate parasitic reflections.
Pt\hkl(-220) and Al$_2$O$_3$\hkl(3-300) have nearly the same lattice spacing and 6-fold symmetry, enabling good epitaxy. This also means that we cannot distinguish between the two peaks in this scan. Au\hkl(-220) has a larger lattice spacing, resulting in the observed shift to the left.
Note that the peak here is primarily limited by the instrumental resolution rather than by the crystallite size, since we opted for a lower angular resolution to achieve a better signal-to-noise ratio. 
For the sputter-deposited sample, only a much broader, less intense Au\hkl(-220) peak is observed, indicating no epitaxy, as we utilize amorphous layers to suppress it. The much lower intensity, as well as the absence of the Pt peak, can be attributed to lower crystal quality already observed in the OOP XRD scans. 

The magnetically important Co layer behaves very differently between the two samples. In the sputter-deposited sample, no Co peak is visible, indicating that the Co IP lattice constant is matched with the surrounding Au, which would indicate that the broad SLD profile in Fig. 1(d) is primarily caused by intermixing and not by roughness or microstructure (grain height variation).
On the other hand, the MBE sample displays a distinct, albeit broad, Co peak, indicating well-defined epitaxy of Co on Au, with no pronounced intermixing from the cap layer. Note that we suspect an hcp order of the Co, which would label the reflex as Co\hkl(2-1-10), but the equivalent fcc Co\hkl(-220) would have the same lattice spacing. We therefore cannot distinguish between the two phases with this measurement.

To further explore the texture, $\phi$-scans were performed on the Au\hkl(-220) peak of both samples, as shown in Fig. 4(b). Both samples show slight intensity variation with sample rotation due to changes in the size of the spot hitting the sample. The sputter-deposited sample has an almost angle-independent intensity, indicative of a polycrystalline structure. Combined with the single fcc\hkl(111) phase visible in the OOP direction and the pole figure, we can deduce a pure fcc\hkl(111) fiber texture for these films, defined by grains, which are well-ordered in the out-of-plane direction, but randomly oriented in the film plane.
On the other hand, the MBE sample shows six sharp peaks, with no intermediate intensity. For an OOP fcc\hkl(111) oriented single crystal, 6 \hkl{-220} equivalent reflections are expected in the film plane. This exact behavior is observed in the $\phi$-scan, indicating good epitaxy on the substrate, as already implied by the pole figure. 

However, the film is not a single large crystal, as shown by the AFM and STM images in Fig. 5. Instead, the film consists of grains whose crystals are all aligned in the same manner. The grains themselves can have different stacking sequences. Even without defects, the grains can nucleate with ABC or CBA stacking, preventing them from merging and resulting in pronounced trenches between the grains. However, we cannot distinguish between these two stacking orders with our X-ray measurements.

\begin{figure}[H]    
    \includegraphics[width=\textwidth]{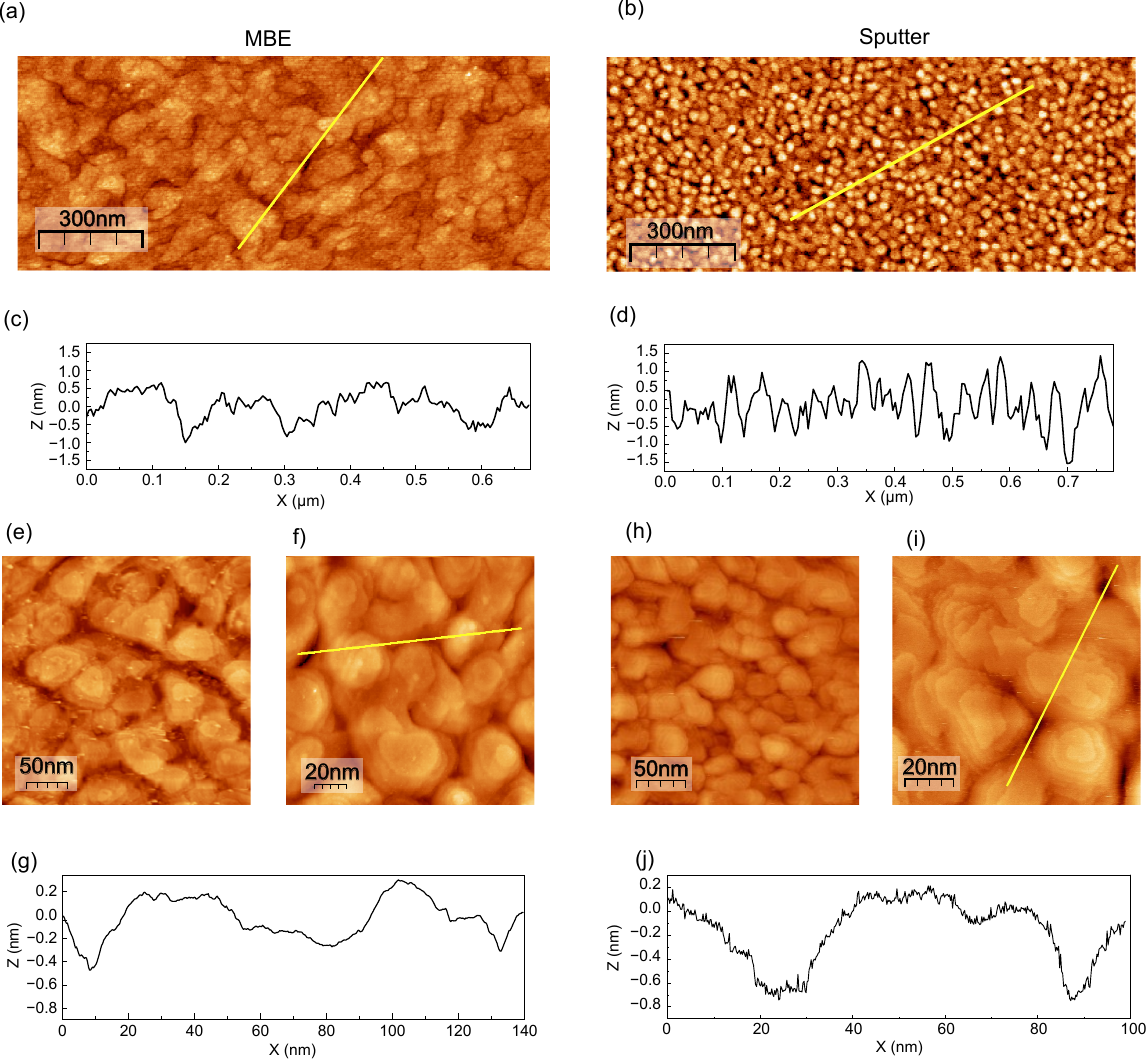}
    \caption{ AFM images of (a) MBE-grown and (b) sputter-deposited samples with their respective line profiles in (c,d); STM images of (e,f) MBE-grown and (h,i) sputter-deposited samples, with their respective line profiles (g,j). The MBE-grown sample shows larger and more laterally interconnected surface features, whereas the sputter-deposited sample exhibits a more isolated granular morphology. Note the slightly different scale bars in (e,h) and (f,i)}
    \label{afm_stm_combined}
\end{figure}

In order to complement the reciprocal space analysis with XRR as well as in- and out-of-plane XRD measurements, we also performed real-space AFM and STM measurements of the surface structure, as displayed in Fig. 5. The large-scale AFM image in Fig. 5(a) reveals that the MBE-grown sample shows a more interconnected terrace-like network, however, still permeated by some quite deep trenches, while the sputter-deposited sample (Fig. 5(b)) consists of a more periodic structure of densely packed, distinct isolated grains. Overall, the height differences of the sputter-deposited sample are within $\pm$\SI{1.5}{\nano\meter} (see Fig. 5(c,d)). However, the MBE-grown sample also reveals a surprisingly large height range of still $\pm$\SI{0.8}{\nano\meter} as extracted from Fig. 5(c).
This still-quite-large surface roughness of the MBE-grown sample can be explained by the presence of crystallites with different Au\hkl(111) stacking sequences, ABC versus CBA, which prevent coalescence of adjacent crystallites with opposite stacking sequences, thus leaving some grain-boundary-like deep trenches between those ABC- and CBA-stacked regions. As a result, the surface roughness and height variations are not markedly different between MBE-grown and sputter-deposited samples. When zooming closer into the surface structure using STM in Fig. 5(e,f,h,i), differences between MBE-grown and sputter-deposited samples become even less pronounced, and for both sample types we observe atomically smooth Au\hkl(111) surface planes on top of terraces and grains with diameters in the range of 20-30\,nm. Here, height differences become quite comparable with $\pm$0.4\,nm for the MBE-grown and $\pm$0.5\,nm for the sputter-deposited sample, as seen in respective line-scans in Fig~5(g,j). Upon careful inspection of the maze-like terrace structure of the MBE-grown samples and the grain structure of the sputter-deposited samples, we noticed that many of the maze-like terraces and grains in both samples exhibit distinct screw dislocations. These screw dislocations show left- as well as right-handedness for both substrate types as can be seen in the supplementary material Fig. S2 for the sputter-deposited sample, while comparable morphologies for MBE-grown samples were reported in reference  \cite{ranaEnantiospecificMagnetoconductanceAsymmetry2026} (Fig. 5(c) as well as its supplementary Fig. S11, S12, S18 and S21). 
\\
We will later see in how far the quite large differences of the two sample types in XRR and XRD analysis, i.e., in reciprocal space, paired with the much less pronounced differences in close up real-space AFM/STM analysis of the surfaces influence the resulting microscopic and macroscopic responses after the adsorption of the peptides on the Au/Co/Au samples, when performing CISS and MIPAC-type experiments.

\subsection{3.2 XPS verification of peptide adsorption on sputter-deposited Au substrates}
\vspace{0.5em}
\begin{figure}[H]
    
    \includegraphics[width=\textwidth]{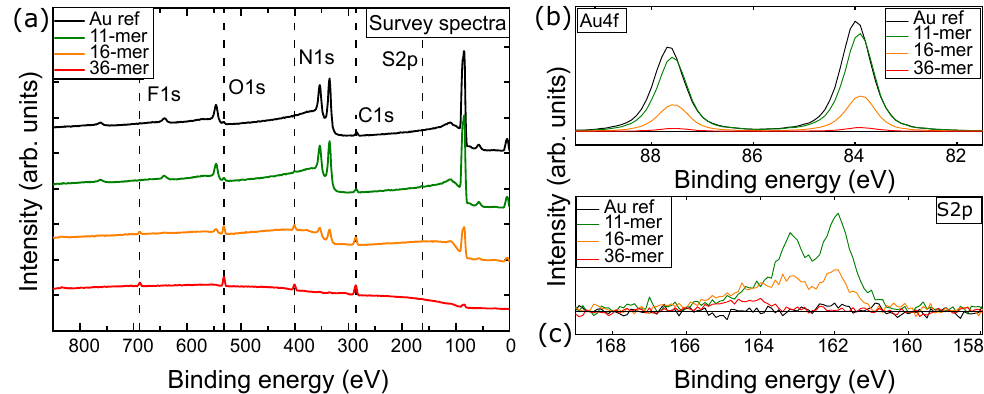}
    \caption{(a) XPS survey spectra of peptides of varying lengths on Au substrates, highlighting N1s, C1s, and O1s core level regions;
    (b) Au4f core level spectra ; (c) S2p core level spectra with respect to the increase in peptide
length. }
    \label{fig:xps_all_combined}
\end{figure}

XPS core level spectra were fitted using a Shirley background for C1s and a quadratic background for S2p, O1s, and N1s. Voigt (and doublet) functions were applied, with literature-reported parameters for the S2p and C1s components \cite{castnerXrayPhotoelectronSpectroscopy1996}.

To investigate the chemical composition and bonding of the peptides to the Au surface via the thiol group, XPS measurements were conducted. Successful deposition of peptides of varying length was confirmed by the appearance and increased intensity of N1s, C1s, and O1s signals (Fig. 6(a)), together with reduced Au4f substrate peaks with increasing peptide length as shown in (Fig. 6(b)). The S2p signal (Fig. 6(c)), characteristic of the cysteine thiol anchor, however, decreases with peptide length and is indistinguishable from noise for the 36-mer. This additionally hints towards the molecules binding to the surface via the sulfur, as the increase in length also increases the layer thickness, thus eventually exceeding the inelastic mean free path of the photoelectrons. Hence, the sulfur can no longer be detected. This is an important finding as it reveals that for long peptide chains, it becomes difficult to find evidence of the cysteine thiol anchoring, as the respective electrons are no longer transmitted through adsorbed molecules. Additionally, deconvolution of the S2p core level reveals two thiol components (161.9 and 163.4\,eV), indicating both Au-bound and free thiols \cite{cavalleriHighResolutionXray2004}, confirming molecular adsorption and suggesting multilayer formation. For the 16-mer, an additional S2p component indicates either contaminants or photoelectron-induced thiol cleavage \cite{cavalleriXPSMeasurementsLcysteine2001}. C1s, N1s, and O1s deconvolution show expected chemical-bond components for peptides \cite{artemenkoReferenceXPSSpectra2021}. Overall, the XPS results confirm successful peptide deposition on the Au surface, while also showing that direct detection of the thiol anchoring signal becomes increasingly difficult for longer peptide chains.

\subsection{3.3 CISS after peptide adsorption (STS I--V and magnetization switching \texorpdfstring{$\Delta z$}{Delta Z})}

\begin{figure}[H]\centering
    
    \includegraphics[width=\textwidth]{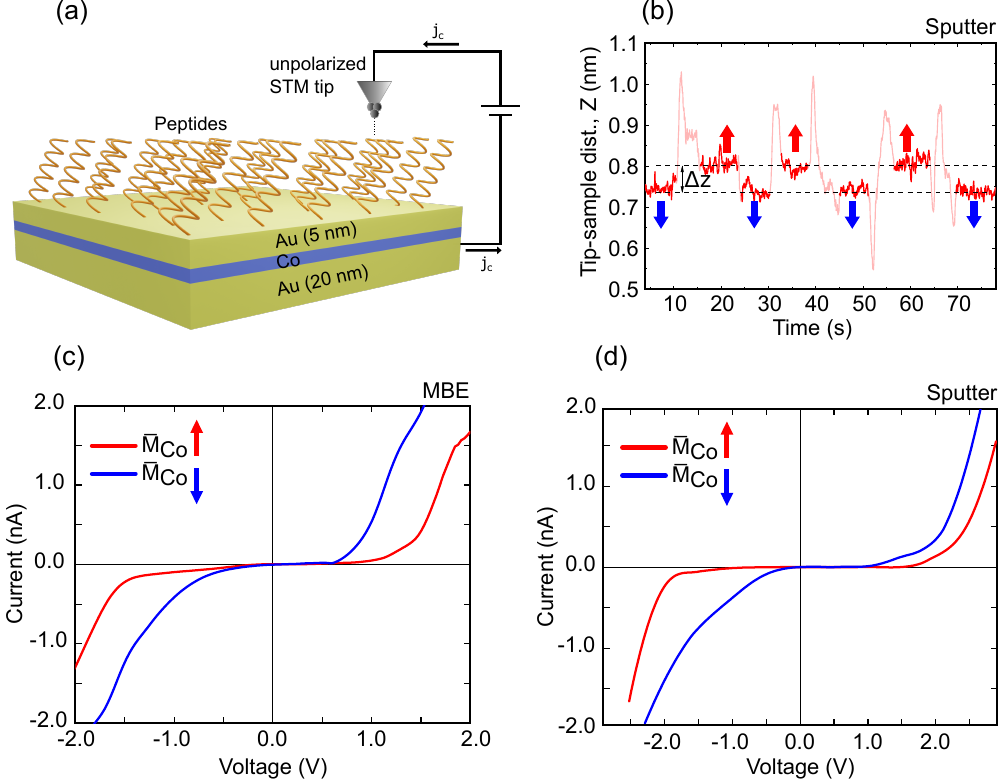}
    \caption{ a) Schematic representation of the STS setup - the Co layer thickness for MBE-grown and sputter-deposited samples are 1.2\,nm and 0.9\,nm, respectively; b) Magnetization-dependent tip-sample distance ($\Delta z$) measurements on the sputter-deposited sample under constant current STM mode; Magnetization-dependent I-V curves measured by STS on c) MBE-grown and d) sputter-deposited  Au/Co/Au films after peptide adsorption, displaying similar differences between up and down magnetization directions.  }
    \label{fig:film_peptide_iv_tip_combined}
\end{figure}

Magnetization-dependent tip-sample distance measurements were performed after peptide exposure (see Fig. 7(a) for illustration), such as previously also reported for MBE-grown samples \cite{nguyenCooperativeEffectElectron2022}. During constant current STM measurements, we alternatingly switched the magnetization direction of the Co layer in-situ up and down using an external magnet and observed the tip retracting upward (for up magnetization) and downward (for down magnetization), as shown in Fig. 7(b). These shifts in tip-sample distance are due to the change in the tunnel magnetoresistance across the peptide/magnetic thin film substrate hybrid spin-valve structure upon magnetization reversal of the Co layer, which is characteristic for the CISS effect, i.e., a well-defined spin polarization occurring at the interface between the peptides and the Au cap layer. 
\\

Additionally, I-V curves were measured by STS on both MBE-grown and sputter-deposited Au/Co/Au samples covered with
the peptides. Measurements for both substrate types (with magnetization up and down) show a systematic and reproducible dependence of tunneling current on the magnetization direction of the Co layer, as shown in Fig. 7(c and d). These results indicate that the electron transport across the sample-peptide interface is sensitive to the chirality of the peptide and the magnetization direction of the Au/Co/Au substrate, thus suggesting a hybrid spin-valve-type giant magnetoresistive (GMR) behaviour of the peptide/ferromagnetic layer composite structure, where the once adsorbed peptides act as the fixed layer and the ferromagnetic Co layer acts as the free layer since its magnetization can be conveniently reversed with an external magnet. These results are consistent with earlier findings \cite{nguyenCooperativeEffectElectron2022}, where similar magnetization-dependent responses were observed in both L- and D-chiral peptides on MBE-grown Au/Co/Au substrates. The spin polarization (SP) was estimated using the equation obtained from literature \cite{nguyenMechanismMolecularCISS2024}:
\begin{equation*}
SP = \frac{I_{\uparrow} - I_{\downarrow}}{I_{\uparrow} + I_{\downarrow}}
\end{equation*}
where $I_{\uparrow}$ and $I_{\downarrow}$ are the tunneling currents measured during up and down Co layer magnetization, respectively. Using this equation, we obtained SP values of 80\% and 77\% for the MBE-grown and sputter-deposited samples, respectively. \\

\subsection{3.4 Macroscopic magnetic response after peptide adsorption (MOKE hysteresis loops and domain-wall dynamics)}

\begin{figure}[H]\centering
    
    \includegraphics[width=\textwidth]{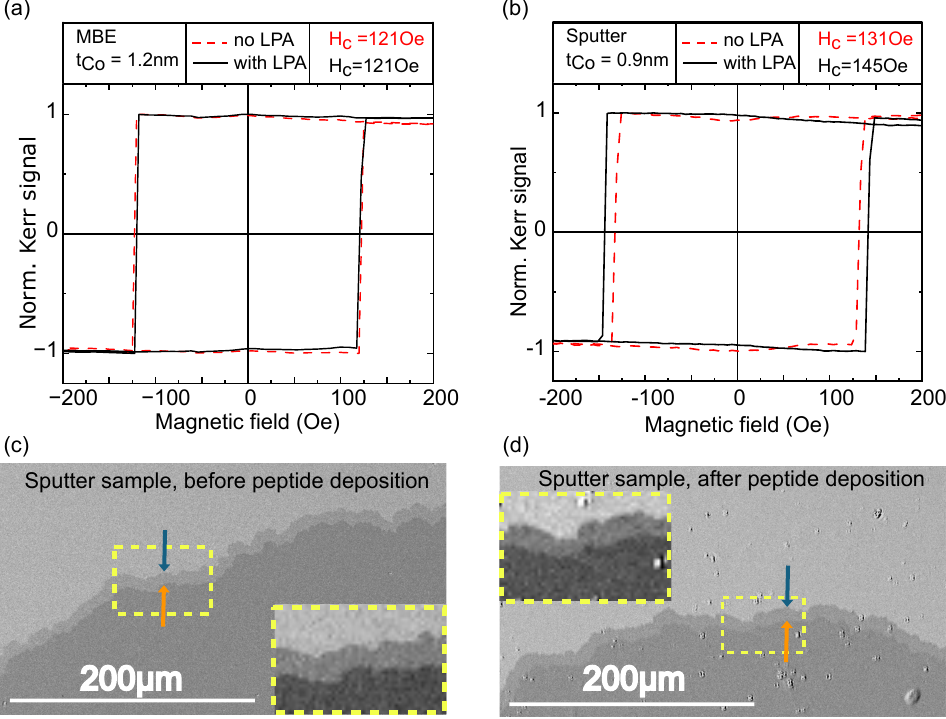}
    \caption{MOKE hysteresis loops before and after peptide exposure (a) MBE-grown Au/Co/Au sample and (b) sputter-deposited Au/Co/Au sample. The sputter-deposited sample has slightly enhanced coercivity after exposure to the peptides, while the MBE-grown sample has no coercivity enhancement after exposure to peptides. Domain wall motion of (c) before and (d) after peptide exposure on the sputter-deposited sample. The orange and blue arrows show the domain wall position before and after the 500\,ms magnetic field pulsing, respectively. }
    \label{fig:moke_hys_combined}

\end{figure}

The hysteresis loops measured using MOKE microscopy show that in the sputter-deposited samples, there is a small but reproducible coercivity enhancement after the peptide exposure. The coercivity increased from 131\,Oe before peptide exposure to 145\,Oe after peptide exposure, i.e., by 11\% as shown in Fig. 8(b). While some small coercivity enhancement upon peptide exposure was detected consistently for all sputter-deposited Co thicknesses with out-of-plane anisotropy, i.e., for Co(0.7\,nm), Co(0.9\,nm), Co(1.0\,nm) thin films (see supplementary material Fig. S5), such a coercivity enhancement was never detected for the MBE-grown samples, as shown for an example in Fig. 8(a). In addition to and consistent with the hysteresis loop measurement of the sputter-deposited sample, it was also observed after exposure to peptides that the velocity of the domain wall under well-defined magnetic field pulses was notably reduced compared to the pristine state. This is shown in Fig. 8(c,d) for the pristine and peptide-exposed sputter-deposited sample, respectively. The domain wall velocity before peptide exposure was $\approx$ \SI{20}{\micro\meter\per\second} and after peptide exposure was reduced to $\approx$ \SI{17}{\micro\meter\per\second}. This reduction in domain wall velocity by about 15\% suggests increased domain-wall pinning at the grain boundaries of the sputter-deposited sample, likely arising from reduced more effective domain wall pinning after peptide exposure \cite{gnoliEnhancementMagneticStability2024,hsuEnhancedMagneticAnisotropy2013,beniniCollapseStandardFerromagnetic2025}. However, in order to determine if this small coercivity enhancement effect is really related to the chirality of the adsorbed peptides, i.e., to a MIPAC effect, more extended experiments using L- and D-chiral peptides as well as similar non-chiral reference molecules need to be performed in the future.\\

\section{4 Conclusions}

The results of this study offer some important insights into whether differences in ferromagnetic substrate properties, such as microstructure, play an important role in observing the CISS and MIPAC-type effects in peptide-magnetic thin film hybrid structures. XRD analysis shows that sputter-deposited Au/Co/Au samples have broader diffraction peaks and less pronounced Laue oscillations as compared to MBE-grown samples, indicating less coherent crystallinity, higher interface roughness and higher density of grain boundaries and surface defects (as shown in the XRR/XRD and AFM/STM characterization). Nevertheless, the CISS effect was observed at the microscopic scale with spin-polarization of 80\% and 77\% for MBE-grown (epitaxial) and sputter-deposited (polycrystalline with a well-defined out-of-plane Au\hkl(111) texture) samples, respectively, as shown by STS measurements and Co layer magnetization-dependent tip–sample distance measurements. 
\\
In contrast to the microscopic CISS studies, the macroscopic magnetic response in MIPAC-type magnetometry experiments seems more sensitive to the structural properties of the samples. The sputter-deposited sample shows, after peptide adsorption, a slightly enhanced coercivity in room temperature MOKE measurements, whereas the MBE-grown samples do not reveal any coercivity enhancement. This suggests that structural features, such as grain isolation via surrounding grain boundaries, interface roughness and surface morphology, can influence how peptide adsorption affects the magnetic reversal process. More specifically, we suggest that in our case, the densely packed, distinct, isolated grains in the sputter-deposited samples are more susceptible to an increased domain wall pinning upon exposure to the peptides and thus to an increased coercivity. In contrast, the MBE-grown samples display a more interconnected terrace network, only occasionally permeated by some quite deep trenches (due to incompatible neighbor grains with ABC versus CBA stacking that cannot coalesce), which is most likely more immune to domain wall pinning. However, if the observed small coercivity enhancement effect in the sputter-deposited sample is really related to the chirality of the adsorbed peptides, i.e., to a MIPAC effect, it still needs to be fully clarified in more specific experiments in the future, where L- and D-chiral versus chemically similar, but non-chiral peptide exposure of sputter-deposited samples needs to be carefully compared. Moreover, the observed symmetric coercivity increase of the sputter-deposited sample for both positive and negative field reversal branches of the hysteresis loop indicates an interaction that most likely does not depend on the chirality of the molecules, as such a chirality sensitive MIPAC interaction  should show a preferential coercivity increase or decrease for one of the two field reversal branches, i.e. the loop should become asymmetric like in an exchange-biased system \cite{raduExchangeBiasEffect2008,noguesExchangeBias1999,maatPerpendicularExchangeBias2001}, rather than only reveal a symmetric coercivity increase or decrease.

\section*{Acknowledgments}

We acknowledge the funding from  Deutsche Forschungsgemeinschaft (DFG) for HYP*MOL TRR-386, TP XX, project number
514664767. \\

We thank Prof. Karin Leistner for providing access to the MOKE microscopy setup.

\bibliographystyle{unsrtnat}
\bibliography{References}

\clearpage

\setcounter{figure}{0}
\setcounter{table}{0}
\renewcommand{\thefigure}{S\arabic{figure}}
\renewcommand{\thetable}{S\arabic{table}}

\begin{center}
{\Large\bfseries Supplementary Information\par}
\vspace{1em}
{\large\bfseries Microscopic and macroscopic characterization:\par}
{\large\bfseries MBE-grown versus sputter-deposited Au/Co/Au thin films\par}
{\large\bfseries for CISS and MIPAC effect studies\par}
\vspace{1em}
{\normalsize Lokesh Rasabathina, Thi Ngoc Ha Nguyen, Aleksandr Kazimir, Rico Ehrler, Julia Krone, Franziska Sch\"olzel, Zihao Liu, Peter Heinig, Markus G\"o\ss ler, Irene Coin, Christina Lamers, Georgeta Salvan, Lech Tomasz Baczewski, Christoph Tegenkamp, and Olav Hellwig\par}
\end{center}

\vspace{1em}
\noindent\rule{\textwidth}{0.4pt}
\vspace{1em}

\section{S1. SQUID-VSM loops for Co thickness series}
\begin{figure}[H]
    
    \includegraphics[width=\textwidth]{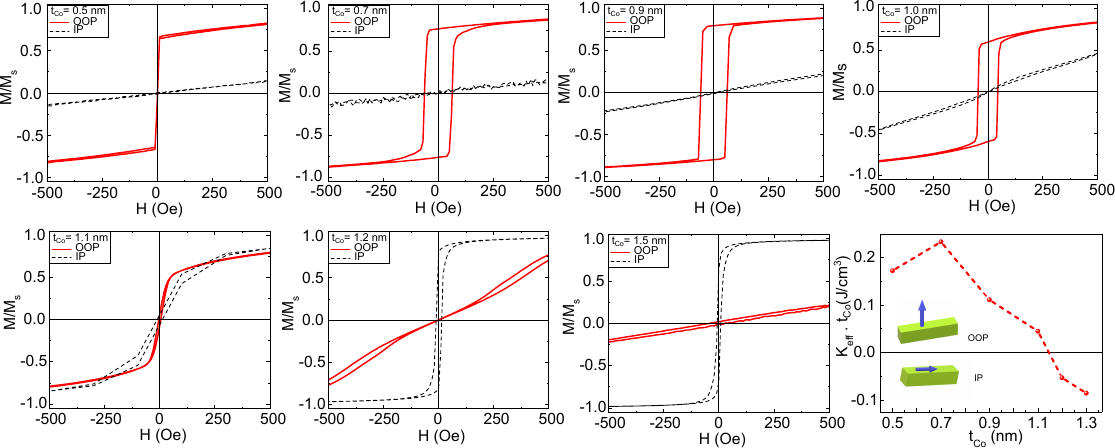}
    \caption{SQUID-VSM hysteresis loops of sputter-deposited thin film samples for various Co thicknesses, displaying the transition from OOP to IP anisotropy as the Co thickness increases.}
    \label{fig:SQUID_series}

\end{figure}
A sputter-deposited Co thickness series was prepared in order to identify a suitable sample with solid perpendicular magnetic anisotropy for the CISS and MIPAC measurements discussed in the main manuscript. The SQUID-VSM hysteresis loops show that the magnetic anisotropy changes systematically as we increase the Co-layer thickness. Based on this thickness-dependent behaviour, the sputter-deposited sample with a Co thickness of 0.9\,nm was selected, since it shows a well-defined out-of-plane hysteresis loop as well as similar coercivity as the MBE-grown sample with a Co thickness of 1.2\,nm. Respective out-of-plane MOKE measurements for a similar MBE-grown Co thickness series can be found in the supplementary material of reference \cite{bendorMagnetizationSwitchingFerromagnets2017} in Fig. SM1.

\section{S2. Coalescence of phase-shifted ABC-stacked crystallites/grains}
\begin{figure}[H]
    
    \includegraphics[width=\textwidth]{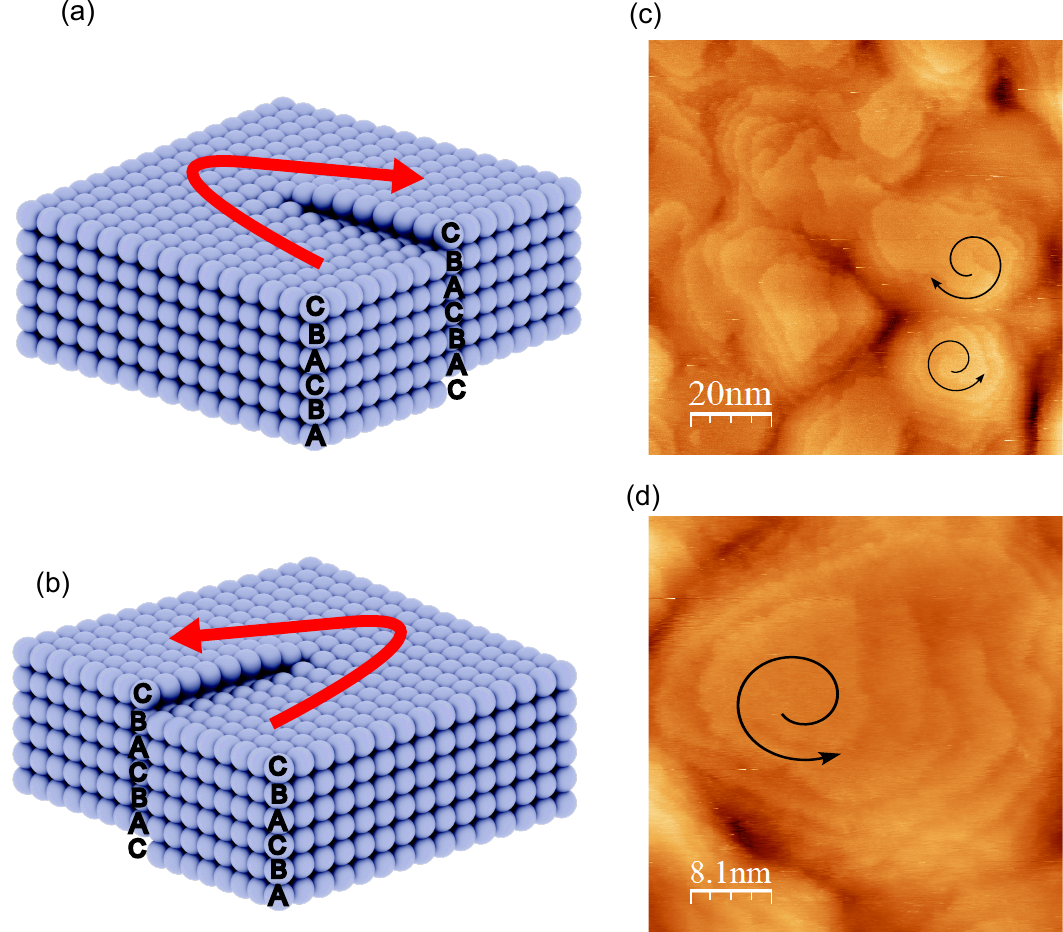}
    \caption{Illustration of screw dislocations for ABC-stacking (a) right-handed and (b) left-handed; (c,d) analogous screw dislocations as observed via STM for the sputter-deposited sample with both left- and right-handedness.}
    \label{abc_stacking}

\end{figure}
The schematics in Fig. S2(a,b) illustrate right- and left-handed screw dislocations for ABC stacking sequence (ABC from bottom to top) occurring in the fcc(111)-oriented Au films for both MBE-grown and sputter-deposited samples. Fig. S2(c,d) show STM images with such right- and left-handed screw dislocations for the sputter-deposited sample. For the MBE-grown samples similar screw dislocations can be found in \cite{ranaEnantiospecificMagnetoconductanceAsymmetry2026}, Fig. 4(c) as well as Supplementary Figs. S11, S12, S18 and S21.
The screw dislocations may originate from coalescing grains or crystallites with the same stacking sequence, but different vertical shift, i.e., for ABC stacking sequence of neighboring grains or crystallites that start with A and B, B and C, or A and C as illustrated in Fig. S2(a,b). For the sputter-deposited samples this can happen for neighboring grains with the same OOP stacking sequence (ABC or CBA) and similar in-plane orientation and for the MBE-grown samples, since it is epitaxial only the stacking sequence of the respective crystallites needs to be identical. Thus, such a coalescence is much more likely for the MBE-grown sample, which then leads to the observed more interconnected terrace-like network,  which however is still permeated by some quite deep trenches due to incompatible ABC- and CBA-stacked neighboring crystallites.
In contrast, the sputter-deposited
sample reveals a more periodic structure of densely packed, distinct isolated grains, with only occasional coalescence.

\begin{figure}[H]
    \centering
    \includegraphics[width=\textwidth]{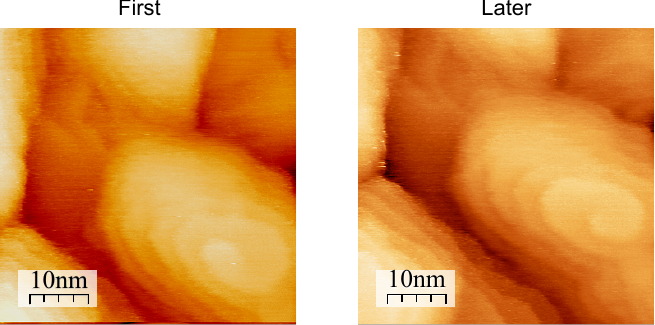}
    \caption{STM images of the same Au terrace region on sputter-deposited sample recorded during the first measurement and again after one day.}
    \label{fig:STM_first_later}
\end{figure}

Repeated STM imaging of the same terrace region after one day shows that the overall terrace morphology remains stable. However, the screw-dislocation-like feature appears to be slightly modified in the later image as shown in Fig. S3. We attribute this local change to mobile or weakly bound Au atoms on the surface, which are able to diffuse on the surface terrace and thus rearrange or move the screw dislocation growth front.

\section{S3. XPS fits}

\begin{figure}[H]
    \centering
    \includegraphics[width=\textwidth]{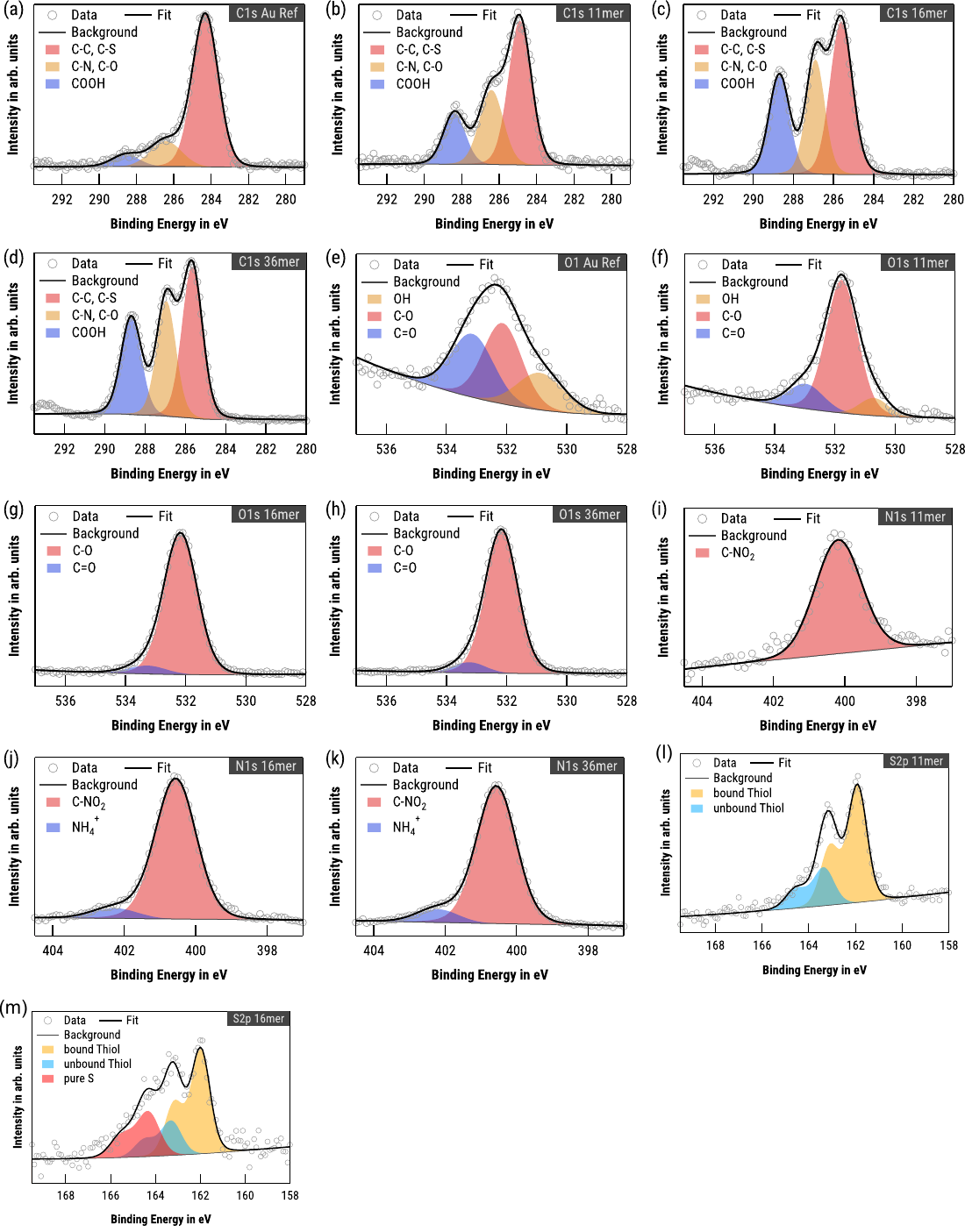}
    \caption{High-resolution XPS core-level spectra and corresponding fits for peptide-exposed Au substrates. The C1s, O1s, N1s, and S2p regions are shown for the Au reference and peptide-covered samples with different peptide lengths. The measured data are shown together with the fitted envelope, background, and individual chemical components used for peak deconvolution. The C1s, O1s, and N1s fits confirm the presence of peptide-related chemical bonds, while the S2p fits show contributions from Au-bound and free thiol species.}
    \label{fig:xps_all_fits_combined}
\end{figure}

The XPS fits support the chemical assignment of the peptide-related signals discussed in the main manuscript. The C1s, O1s, and N1s regions show the expected chemical-bond components associated with the peptide backbone and side groups. In the S2p region, the fitted components indicate the presence of both Au-bound and non-bound thiol species, consistent with cysteine-based attachment to the Au surface. The S2p signal becomes weaker with increasing peptide length, which is consistent with attenuation of the sulfur signal from the Au–S interface by the thicker molecular layer.

\section{S4. Coercivity enhancement studies upon peptide exposure using MOKE magnetometry}
\begin{figure}[H]
    \centering
    \includegraphics[width=0.50\textwidth]{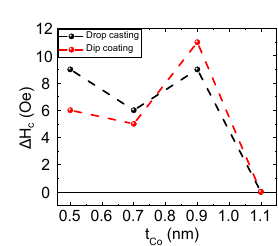}
    \caption{Coercivity enhancement measured by MOKE magnetometry for sputter-deposited samples with different Co thicknesses.}
    \label{fig:moke_coercivity_series}
\end{figure}
MOKE hysteresis loops were measured before and after peptide deposition for Co samples with 0.5\,nm, 0.7\,nm, 0.9\,nm and 1.1\,nm. We observed coercivity enhancements for all Co thicknesses in the range between 5 and 12\,Oe, except for the almost already in-plane easy axis Co 1.1 nm sample, where no significant coercivity enhancement was observed.

\end{document}